\documentstyle[pre,aps,multicol,eqsecnum,epsf,rotate]{revtex}

\begin{document}
\draft
\tighten


\title{\LARGE Interactions, Distribution of Pinning Energies, and Transport in
	      the Bose Glass Phase of Vortices in Superconductors}

\author{Uwe C. T\"auber and David R. Nelson}

\address{Lyman Laboratory of Physics, Harvard University, Cambridge,
         Massachusetts 02138}

\date{\today}
\maketitle


\begin{abstract}

We study the ground state and low energy excitations of vortices pinned to
columnar defects in superconductors, taking into account the long--range
interaction between the fluxons. We consider the ``underfilled'' situation in
the Bose glass phase, where each flux line is attached to one of the defects,
while some pins remain unoccupied. By exploiting an analogy with disordered
semiconductors, we calculate the spatial configurations in the ground state, as
well as the distribution of pinning energies, using a zero--temperature Monte
Carlo algorithm minimizing the total energy with respect to all possible
one--vortex transfers. Intervortex repulsion leads to strong correlations
whenever the London penetration depth exceeds the fluxon spacing. A pronounced
peak appears in the static structure factor $S(q)$ for low filling fractions
$f \leq 0.3$. Interactions lead to a broad Coulomb gap in the distribution of
pinning energies $g(\epsilon)$ near the chemical potential $\mu$, separating
the occupied and empty pins. The vanishing of $g(\epsilon)$ at $\mu$ leads to a
considerable reduction of variable--range hopping vortex transport by
correlated flux line pinning.

\end{abstract}


\begin{multicols}{2}


\section{Introduction}
 \label{introd}

Recently, the static and dynamic properties of flux lines in high--temperature
superconductors subject to an external magnetic field $\bf B$ have been
intensively studied both experimentally and theoretically \cite{review}. From a
theoretical point of view, the importance of thermal fluctuations \cite{nelson}
and the strong influence of point defects \cite{larkin,fisher} pose a variety
of interesting problems, leading to strikingly rich and complex phase diagrams
\cite{alamos}. For the purpose of applying (high--temperature) superconductors
in external magnetic fields, an effective vortex pinning mechanism is
essential, in order to minimize dissipative losses caused by the Lorentz--force
induced movement of flux lines across the sample. Therefore, in addition to
point defects, the influence of extended or {\it correlated} disorder,
promising stronger pinning effects, on vortex transport properties has been
considered. Experimentally, it has been found that by high--energy ion
irradiation linear damage tracks are formed in the material. These columnar
pins bind the flux lines very strongly, thus significantly increasing the
critical current and shifting the irreversibility line upwards \cite{colpin}.
On the theoretical side, the pinning of flux lines to linear defects and the
ensuing transport properties have been studied by Lyuksyutov, \cite{lyutov} and
in detail by Nelson and Vinokur \cite{nelvin}. In Ref.~\cite{nelvin},
correlated pinning by twin and grain boundaries is discussed as well, a topic
which has been more thoroughly investigated recently by Marchetti and Vinokur
\cite{marvin}.

Similar to the physics of flux lines in pure systems, \cite{nelson} the
statistical mechanics of vortices interacting with columnar pinning centers,
which are aligned parallel to the magnetic field (Fig.~\ref{colpin}), may be
mapped onto the quantum mechanics of bosons in two dimensions \cite{nelvin}
(see Sec.~\ref{modeqs}). At high temperatures, one finds an entangled liquid of
delocalized vortices, separated by a sharp second--order phase transition
(corresponding to a boson localization transition \cite{disbos}) from a
low--temperature phase of lines strongly attached to the columnar pins. This
Bose glass phase is characterized by an infinite tilt modulus, and was shown to
be stable over a certain range of tipping angles of ${\bf B}$ away from the
direction parallel to the linear pinning centers (transverse Meissner effect)
\cite{nelvin}. In addition, the theory also suggests a Mott insulator phase at
low temperatures, with both tilt and compressional moduli acquiring infinite
values, occurring when the vortex density matches the density of columnar pins
\cite{nelvin}. The ensuing phase diagram is sketched schematically in
Fig.~\ref{phadia}. Since the Mott insulator phase is predicted to occur at low
temperatures deep inside a regime with very slow dynamics, it may actually be
difficult to access for kinetic reasons. It is also possible that the Mott
insulator vanishes entirely for sufficiently strong disorder. Some of the
predictions of Ref.~\cite{nelvin} have now been tested and confirmed by
experiments on a number of different samples
\cite{locexp,stmexp,decexp,trexp1,trexp2,trexp3}. E.g., the scaling behavior
near the boson localization transition was studied, \cite{locexp} and in
addition even some properties of the Bose glass itself have been investigated
to some detail \cite{trexp1,trexp2,trexp3}.

In the Bose glass phase, the linear resistivity vanishes for low external
currents $J \ll J_c$, and the most important mechanism for vortex transport is
``tunneling'' between different columnar defect sites via the formation of a
pair of ``superkinks'' \cite{nelvin}, i.e.: a fluxon forms a tonguelike double
kink extending from one pin to another, usually a distant defect with nearly
the same energy, such that the tunneling probability between the columnar
defects is optimized (Fig.~\ref{vrkink}). This is very closely related to
variable--range hopping transport of charge carriers in disordered
semiconductors, \cite{cgapth} and leads to the highly nonlinear expression
\cite{kbnote}
\begin{equation}
	{\cal E} = \rho_0 J \exp \left[ - (E_{\rm K} / T) (J_0 / J)^p \right]
 \label{ivchar}
\end{equation}
for the current--induced electric field ${\cal E}(J)$ in the limit of very low
currents $J \ll J_c$ (Fig.~\ref{curvol}). In Eq.~(\ref{ivchar}), $\rho_0$
denotes the normal--state resistivity, $E_K$ is a typical kink energy, and
$J_0$ sets the current scale, see Eq.~(\ref{curscl}) and Sec.~\ref{ivcvrh}. By
assuming a short--range repulsive interaction between the flux lines, Nelson
and Vinokur could identify $p$ with the Mott variable--range--hopping exponent
in two dimensions, $p = 1/3$.

Similarly, for the case of parallel planar defects containing the ${\bf B}$
axis, Marchetti and Vinokur have found an analagous formula, with $p = 1/2$,
demonstrating that twin and grain boundaries serve as even better pinning
centers for flux lines \cite{marvin}. Returning to linear defects, Hwa,
Le Doussal, Nelson, and Vinokur have suggested that an even more drastic
reduction of vortex motion may be obtained by introducing a controlled splay,
i.e., dispersion in the orientation of the columnar pins \cite{splays}. A new
low--temperature phase, the ``splayed glass'', is found which is related to the
Bose glass, but has a finite equilibrium tilt modulus (however, the system is
dynamically frozen due to diverging energy barriers preventing the relaxation
of small externally induced tilts). Flux--line transport in this phase should
be more strongly inhibited than in the Bose glass for two reasons: (i)
variable--range hopping must now proceed via a much slower process of selecting
both energy {\it and} angle; as a consequence, the exponent in the nonlinear
current--voltage characteristics (\ref{ivchar}) is found to be enhanced to
$p = 3/5$ \cite{splays}. (ii) Flux lines in the splayed glass ground state are
entangled and can only move by cutting through each other, at the expense of a
considerable amount of (free) energy \cite{carloc}.

In the above studies, \cite{nelvin,marvin,splays} the intervortex repulsion was
only treated approximatively, using order--of--magnitude estimates.
Specifically, it was assumed as being essentially short--range. However, in
fact the interaction range is set by the London penetration depth $\lambda$,
which diverges at the critical temperature, and thus the intervortex repulsion
may effectively become very long--range, namely when $\lambda$ is large
compared to the average spacing between flux lines,
$\lambda \gg a_0 = (4/3)^{1/4} (\phi_0 / B)^{1/2}$ (here $\phi_0 = h c / 2 e$
denotes the magnetic flux quantum). These strong long--range interactions will
then determine the width and shape of the distribution of pinning energies,
which, as has been emphasized by Gurevich, \cite{gurevi} in turn strongly
influences vortex transport properties \cite{nelvin}.

Simultaneous observations of both columnar defects and magnetic flux lines,
using scanneling tunneling microscopy, have recently been reported for
${\rm NbSe_2}$ samples in the ``underfilled'' regime, where the number of
columnar pins $N_{\rm D}$ exceeds the number of vortices $N$ \cite{stmexp}.
For the high--temperature superconducting material
${\rm Bi_2Sr_2CaCu_2O_{8+\delta}}$ (BSCCO), a combined chemical etching and
magnetic decoration technique has been similarly successful in determining the
positions of both the columnar defects and vortices; and indeed cases have been
found, in which the strongly correlated spatial distribution of the flux lines
suggests that the intervortex repulsion clearly dominates in magnitude over the
unavoidable fluctuations in pinning energies stemming from the variations in
the ion track diameters \cite{decexp}. This would already imply a noticeable
reduction of vortex transport; for the width of the distribution function
$g(\epsilon)$ of the vortex pinning energies $\epsilon$, is then no longer
determined by the comparatively small variations in pin diameters, but by the
much larger typical interaction scale. Thus the value of $g(\epsilon)$ at the
chemical potential has to be reduced, leading to a greater value for the
current scale in Eq.~(\ref{ivchar}),
\begin{equation}
	J_0 = c / \phi_0 g(\mu) d^3 \quad ,
 \label{curscl}
\end{equation}
where $d$ is the average defect distance. Therefore, by a suitable
``tailoring'' of $g(\epsilon)$, the prefactor of the power law in the exponent
of the current--voltage characteristics (\ref{ivchar}) may be adjusted in order
to achieve more effective pinning.

Moreover, by further exploiting the analogy to two--dimensional bosons
localized at randomly distributed defect sites, one has to at least consider
the possibility that these spatial correlations may produce a ``Coulomb'' gap
\cite{cgapth} in the distribution of pinning energies $g(\epsilon)$ near the
chemical potential $\mu$, which separates the filled and empty energy levels.
In the limit of infinitely long--range interactions,
$\lambda \rightarrow \infty$, one would expect $g(\epsilon)$ to vanish near
$\mu$ according to a power law
\begin{equation}
	g(\epsilon) \propto | \epsilon - \mu |^s \quad ,
 \label{gapexp}
\end{equation}
see Sec.~\ref{dosgap}. This of course invalidates Eq.~(\ref{ivchar}), because
there a finite value of $g(\mu) \not= 0$ has been implicitly assumed
\cite{nelvin}. Instead, a simple re--analysis of the optimization procedure
leading to Eq.~(\ref{ivchar}) shows that the exponent $p_0 = 1/3$ is to be
replaced by the {\it larger} value $p = (s + 1) / (s + 3)$, thus strongly
suppressing vortex motion (see Sec.~\ref{ivcvrh}; also the current scale $J_0$
becomes modified).

In order to determine the gap exponent $s$, and thus $p$, and also to
understand down to which values of $\lambda / d$ and $a_0 / d$ the intervortex
repulsion may still be considered as sufficiently long--range to produce a
correlation--induced Coulomb gap, one has to resort to numerical simulations in
the spirit of previous investigations of the so--called Coulomb glass
(localized charge carriers interacting with a $1/r$ potential)
\cite{leesim,recsim,gapsim}. Except for a mean--field analysis for the Coulomb
potential problem, \cite{cgapth} the only related prior investigation appears
to be a scaling analysis of logarithmically interacting vortices in two
dimensions by Fisher, Tokuyasu, and Young \cite{ftyexp}. We have thus performed
an extensive Monte--Carlo study of a two--dimensional system of $N$ localized
(i.e., classical) particles subject to a Bessel function potential,
representing the interacting flux lines, as a function of the two parameters
$\lambda / d$ and $f = N / N_{\rm D} = d^2 / a_0^2$. We find that for fillings
$0.1 \leq f \leq 0.4$ ($d \approx a_0$ in this range), and large values
$\lambda / d \geq 5$ a very wide and pronounced gap appears in the distribution
of pinning energies, described by an effective gap exponent assuming values up
to $s_{\rm eff} \approx 3$, and as a consequence the transport exponent reaches
at maximum $p_{\rm eff} \approx 2/3$. This means that correlation effects due
to long--range repulsive forces may drastically enhance the correlated pinning
of flux lines in the Bose glass phase. However, even for the case
$\lambda \approx d \approx a_0$, one still finds a marked minimum in
$g(\epsilon)$, with an effective gap exponent $s_{\rm eff} \approx 1$, and
hence $p_{\rm eff} \approx 1/2$. Only for $\lambda / d \leq 0.5$ is the
non--interacting result $p_0 = 1/3$ recovered. We conclude that the correlation
effects due to the intervortex repulsion are surprisingly strong, and may
actually be an interesting means to ``tailor'' the distribution of pinning
energies in ion--irradiated samples, such that flux creep is suppressed as
effectively as possible. A preliminary account of some of these results and
their connection with recent experiments \cite{decexp} has been given in
Ref.~\cite{letter}.

We remark that we do not expect that additional point defects (e.g., oxygen
vacancies) could alter our results substantially. First, pinning to point
defects is subject to much stronger thermal renormalization than pinning to
correlated disorder. Second, for very low currents, these point defects might
indeed trap the spreading of the double--kink configuration considered above;
however, this becomes effective only on unphysically large length scales (in
the km range), \cite{nelvin} and thus is irrelevant for realistic samples.
Finally, no strong pinning effects have been reported in superconducting
materials {\it above} the irreversibility line prior to heavy--ion irradiation.

This paper is organized as follows. In the subsequent section we describe how
our effective Hamiltonian may be derived from the free energy of interacting
flux lines subject to columnar disorder, introduce the relevant parameters and
energy scales, and discuss the region in the phase diagram where we expect our
model to be valid. Furthermore, we describe our specific Monte Carlo simulation
algorithm, also discussing the different possible choices of initial
configurations for the energy minimization procedure. In Sec.~\ref{result}, we
describe our numerical results and give a number of examples for the spatial
configurations we find, as well as for the shape of the ensuing distribution of
pinning energies, as we vary the filling $f$ and the interaction range
$\lambda / d$. We also discuss vortex transport mechanisms in the Bose glass
phase, and deduce the form of the current--voltage characteristics in the
variable--range hopping regime from the previously determined distribution of
pinning energies, identifying the effective Mott exponent $p_{\rm eff}$. A
brief summary and discussion of our results concludes this work.

\section{Description of the Model and Monte Carlo Simulation Procedure}
 \label{modsim}


\subsection{Model equations}
 \label{modeqs}

We start by considering the following model free energy for $N$ flux lines,
defined by their trajectories ${\bf r}_i(z)$ as they traverse the
superconducting sample of thickness $L$, interacting with each other and with
$N_{\rm D}$ columnar defects, parallel to the magnetic field ${\bf B}$ which is
aligned along the $z$ axis (perpendicular to the copper oxide planes in the
case of the anisotropic high--$T_c$ materials), \cite{nelvin} (see
Fig.~\ref{colpin})
\begin{eqnarray}
  F_N[\{ {\bf r}_i \}] = \int_0^L dz \sum_{i=1}^N &&\Biggl\{
    {{\tilde \epsilon}_1 \over 2} \left( {d {\bf r}_i(z) \over dz} \right)^2 +
    {1 \over 2} \sum_{j \not= i}^N V[r_{ij}(z)] \nonumber \\ &&+
    \sum_{k=1}^{N_{\rm D}} V_{\rm D} \left[ {\bf r}_i(z) - {\bf R}_k(z) \right]
	\Biggr\} \; .
 \label{modelf}
\end{eqnarray}
The first term describes the elastic line tension, and stems from an expansion
with respect to small tipping angles of the line energy of nearly straight
vortices \cite{nelson}. For a dense liquid of flux lines, $\lambda \gg a_0$,
the tilt modulus is ${\tilde \epsilon}_1 \approx (M_\perp / M_z) \epsilon_0
\ln (a_0 / \xi)$, where the material anisotropy is embodied in the
effective--mass ratio $M_\perp / M_z$ ($\ll 1$ for high--temperature
superconductors), $\xi$ is the (in--plane) coherent length
($\kappa = \lambda / \xi \approx 100$), and
$\epsilon_0 = (\phi_0 / 4 \pi \lambda)^2$ sets the energy scale. On the other
hand, for low fields $B \leq \phi_0 / \lambda^2$, i.e.: $\lambda \leq a_0$,
${\tilde \epsilon}_1 \approx \epsilon_0$ because of the magnetic coupling
between the copper oxide planes \cite{fisher}. In this situation the vortex
lines are therefore considerably less flexible than at high fields.

Furthermore, $r_{ij}(z) = | {\bf r}_i(z) - {\bf r}_j(z) |$, and
\begin{equation}
	V(r) = 2 \epsilon_0 K_0 \left( r / \lambda \right)
 \label{bespot}
\end{equation}
represents the repulsive interaction potential between the lines; the modified
Bessel function $K_0(r / \lambda)$ describes a screened logarithmic
interaction, for $K_0(x) \propto - \ln x$ for $x \rightarrow 0$, and
$K_0(x) \propto x^{-1/2} e^{-x}$ for $x \rightarrow \infty$. Thus the
(in--plane) London penetration depth $\lambda$ defines the interaction range.
Note that according to the two--fluid model, the penetration depth diverges at
the zero--field transition temperature $T_c$,
\begin{equation}
	\lambda(T) = \lambda_0 \left[ 1 - (T/T_c)^4 \right]^{-1/2} \quad ,
 \label{london}
\end{equation}
and therefore the interaction range becomes considerably longer upon
approaching $T_c$; e.g., for $T / T_c \approx 0.93$ one has
$\lambda(T) / \lambda_0 = 2$. Because $\epsilon_0 \propto \lambda^{-2}$, the
interactions are both weak and long--range near $T_c$.

Finally, the columnar pins are described by a sum of $N_{\rm D}$
$z$--independent potential wells $V_{\rm D}$, with average spacing $d$,
centered on the randomly distributed positions $\{ {\bf R}_k \}$. The damage
track diameters, produced by heavy--ion irradiation, are typically
$2 c_0 \approx 100 \AA$, with a variation induced by the root mean--square
dispersion of the ion beam of $\delta c_k / c_0 \approx 15 \%$. The pinning
energies $U_k$ are related to the column diameters via the (interpolation)
formula \cite{nelvin}
\begin{equation}
	U_k \approx {\epsilon_0 \over 2}
                    \ln \left[ 1 + ( c_k / \sqrt{2} \xi)^2 \right] \quad ,
 \label{pinpot}
\end{equation}
the variation of which thus induces a certain distribution of the pinning
energies $P$, with a width $w = \sqrt{\langle \delta U_k^2 \rangle}$ determined
from Eq.~(\ref{pinpot}),
\begin{equation}
	w = {\epsilon_0 \over 1 + (\sqrt{2} \xi / c_0)^2}
	    {\delta c_k \over c_0} \quad .
 \label{varpot}
\end{equation}
(Notice that here the repulsive vortex interactions have not been accounted
for.) Above a certain temperature $T_0$, however, the effective radius of the
normal--conducting region at the defects will be given by the coherence length,
which grows with temperature according to
\begin{equation}
	\xi(T) = \xi_0 (1 - T/T_c)^{-1/2} \quad ;
 \label{cohlen}
\end{equation}
hence $\sqrt{2} \xi(T)$ should be inserted into Eq.~(\ref{pinpot}) instead of
$c_k$, whenever $\sqrt{2} \xi(T) \geq c_0$, or $T \geq T_0$, with
\begin{equation}
	T_0 = T_c \left( 1 - 2 \xi_0^2 / c_0^2 \right) \quad .
 \label{effpin}
\end{equation}
Consequently, above $T_0$ the effective fluctuations of the pinning energies
$P$ (without interactions) will be smoothened out. For $c_0 \approx 50 \AA$ and
$\xi_0 \approx 10 \AA$, e.g., this will happen at $T_0 \approx 0.92 T_c$, and
the characteristic pinning energies  will be
$U_0 = \langle U_k \rangle \approx 0.35 \epsilon_0$. On the other hand, at
$T = 0$ one finds instead $U_0 \approx 1.3 \epsilon_0$, and, assuming
$\delta c_k / c_k \approx 0.15$ one gets $w \approx 0.13 \epsilon_0$, while for
$T = 0.8 T_c$ one has $\xi \approx 2.24 \xi_0$, leading to
$U_0 \approx 0.63 \epsilon_0$ and $w \approx 0.11 \epsilon_0$.

The thermodynamic properties of the model (\ref{modelf}) may be derived from
the grand--canonical partition function \cite{nelvin}
\begin{equation}
	{\cal Z}_{\rm gr} = \sum_{N = 0}^\infty {1 \over N !} e^{\mu N L / T}
	\int \prod_{i = 1}^N {\cal D}[{\bf r}_i(z)]
	e^{- F_N[\{ {\bf r}_i(z) \}] / T} \quad ,
 \label{grpart}
\end{equation}
where $\mu$ denotes the chemical potential
\begin{equation}
	\mu = \epsilon_0 \ln \kappa - {\phi_0 \over 4 \pi} H
            = {\phi_0 \over 4 \pi} \left( H_{c_1} - H \right) \quad ,
 \label{chemmu}
\end{equation}
which changes sign at the lower critical field
$H_{c_1} = \phi_0 \ln \kappa / 4 \pi \lambda^2$. As is explained in detail in
Refs.~\cite{nelson,nelvin}, the statistical mechanics of (\ref{modelf}) and
(\ref{grpart}) can be formally mapped onto a two--dimensional quantum
mechanical problem by using a transfer matrix approach, the physical $z$
direction being interpreted as an imaginary time axis. In the thermodynamic
limit $L \rightarrow \infty$, the properties of our model (\ref{modelf}) are
determined by the lowest energy eigenvalue, i.e., the corresponding
ground--state wave function, which is symmetric with respect to exchange of
flux lines, and thus of bosonic character; furthermore the effective
temperature of this system of interacting bosons is then to be taken as zero.
The mapping of the vortex line problem to the quantum mechanics of charged
bosons is summarized in Table~\ref{bosmap} (from Ref.~\cite{nelvin}). Notice
that in this analogy, the real temperature $T$ assumes the role of Planck's
constant $\hbar$ in the quantum representation, and the boson electric field
and current density map on the superconducting current $J$ and the true
electric field ${\cal E}$, respectively; hence the roles of conductivity and
resistivity become interchanged.

In the following, we shall be interested in the low--temperature properties of
flux lines in the Bose glass phase, pinned to columnar defects, with filling
fraction
\begin{equation}
	f = N / N_{\rm D} = (d / a_0)^2 = B / B_\phi < 1 \quad .
 \label{fillin}
\end{equation}
More specifically, we need $B < B^*(T)$, where vortex interactions become
important in determining the localization length, and thus affect the pinning
of the flux lines to the columnar defects considerably \cite{nelvin}. In the
temperature range of interest here,
\begin{equation}
	B^*(T) \approx \left\{
	\begin{array}{ll}
              B_\phi = (4/3)^{1/2} \phi_0/d^2      & \mbox{for $T < T_0$} \\
	      B_\phi (c_0 / 2 \xi_0)^2 (1 - T/T_c) & \mbox{for $T_0 < T < T_1$}
 	\quad . \end{array} \right.
 \label{crossb}
\end{equation}
$T_1$ denotes the temperature above which the entropy associated with thermal
fluctuations becomes relevant for the vortex pinning, by modifying the
localization length and thus renormalizing the binding free energies (for the
temperature dependence of $B^*(T)$ for $T > T_1$, see Ref.~\cite{nelvin},
App.~D). The estimate of Ref.~\cite{nelvin} is
\begin{equation}
     T_1 \approx T_c {(c_0 / 4 \xi_0) (\ln \kappa / {\rm Gi})^{1/2} \over
                  1 + (c_0 / 4 \xi_0) (\ln \kappa / {\rm Gi})^{1/2}} \quad ,
 \label{crosst}
\end{equation}
where ${\rm Gi} = (M_z / 2 M_\perp) (T_c / H_c^2 \xi_0^3)^2$ is the so--called
Ginzburg number, and the thermodynamic critical field at zero temperature is
$H_c = \sqrt{2} \phi_0 / 4 \pi \lambda_0 \xi_0$. For YBCO, typically
${\rm Gi} \approx 0.01$, and with $\kappa \approx 100$ we get
$T_1 / T_c \approx 0.96$ and $B^*(T_1) \approx 0.25 B_\phi$; for the highly
anisotropic material BSCCO, upon using ${\rm Gi} \approx 0.1$ instead, one
finds $T_1 / T_c \approx 0.89$ and $B^*(T_1) \approx 0.69 B_\phi$. For
$T < T_1$, each flux line is localized on essentially one columnar defect, and
thermal fluctuations play a very minor role; thus all temperatures in this
range may be considered as ``low''. At $T_1$, entropic effects become
important, and at the even higher so--called depinning temperature $T_{\rm dp}$
the localization length of a single vortex is determined by the interplay of a
large number of pins (see Fig.~\ref{phadia}). Similarly to $T_1$, $T_{\rm dp}$
may be estimated as \cite{nelvin}
\begin{equation}
T_{\rm dp} \approx T_c {(c_0 \alpha / 4 \xi_0) (\ln \kappa / {\rm Gi})^{1/2}
        \over 1 + (c_0 \alpha / 4 \xi_0) (\ln \kappa / {\rm Gi})^{1/2}} \quad ,
 \label{deptem}
\end{equation}
where $\alpha = \ln^{1/2} [(d / \sqrt{2} \xi_0) (1 - T_1/T_c)^{1/2} ]$. With
$d \approx 1000 \AA$, e.g., using the above numbers one finds
$T_{\rm dp}/T_c \approx 0.98$ for YBCO, and $T_{\rm dp}/T_c \approx 0.94$ for
BSCCO.

For $T$ less than the characteristic fluctuation temperature $T_1$, we may thus
simply consider the classical limit of the corresponding boson problem
($\hbar \rightarrow 0$); furthermore, as the vortices are well separated in
this regime [with $B < B^*(T)$], the Bose statistics become irrelevant. Thermal
wandering is effectively suppressed, and the flux lines will be essentially
straight; hence the tilt energy in Eq.~(\ref{modelf}) can be neglected.
Therefore, in the boson representation we eventually have to consider the
time--independent problem of $N$ interacting particles located at
$N_{\rm D} > N$ available defect positions. Thus we define our model by the
two--dimensional effective Hamiltonian
\begin{equation}
	H = {1 \over 2} \sum_{i \not= j}^{N_{\rm D}} n_i n_j V(r_{ij})
                      + \sum_{i = 1}^{N_{\rm D}} n_i t_i \quad ,
 \label{effham}
\end{equation}
and its grand--canonical counterpart
\begin{equation}
	{\tilde H} = H - \mu \sum_{i = 1}^{N_{\rm D}} n_i \quad .
 \label{grcham}
\end{equation}
Here $i,j = 1,\ldots,N_{\rm D}$ denote the defect sites, randomly distributed
on the $xy$ plane; $n_i = 0, 1$ represents the corresponding site occupation
number ($n_i = 1$ if a flux line is bound to columnar defect $i$),
$\sum_{i=1}^{N_{\rm D}} n_i = N$. We have also included random site energies
$t_i$, originating in the variation of pin diameters [see Eq.~(\ref{varpot})].
Their distribution $P$ can be chosen to be centered at $\langle t \rangle = 0$,
with width $w$; this amounts to absorbing the average pinning energy
$U_0 = \langle U_k \rangle$ (which may include small thermal renormalizations)
into the chemical potential $\mu$. Realistically, the probability distribution
of the site energies would likely be Gaussian; \cite{decexp} however, for
simplicity we shall assume a flat distribution
\begin{equation}
	P(t_i) = \Theta (w - |t_i|) / 2w
 \label{sedist}
\end{equation}
[$\Theta(x)$ denotes the Heaviside step function]. As the total shape and width
of the distribution of pinning energies will turn out to be determined by the
interactions, the precise form of $P(t_i)$ is of minor importance. Note that
$P$ corresponds to a bare density of states.

For the {\it interacting} system (\ref{effham}), we define single--particle
site energies $\epsilon_i$ as follows: For filled sites ($n_i = 1$),
$\epsilon_i$ is the energy required to remove the particle at site $i$ to
infinity; for empty sites ($n_i = 0$), correspondingly $\epsilon_i$ is the
energy needed to introduce an {\it additional} particle from infinity to site
$i$. Equivalently, we may state that $\epsilon_i$ constitutes the potential
energy on site $i$, resulting from the disorder {\it and} the interaction with
all other occupied sites $j$, or
\begin{equation}
	\epsilon_i = {\partial H \over \partial n_i}
                   = \sum_{j \not= i}^{N_{\rm D}} n_j V(r_{ij}) + t_i \quad .
 \label{epsili}
\end{equation}
In thermal equilibrium, the chemical potential $\mu$ separates the occupied and
empty states, $\epsilon_i \leq \mu$ for all $i$ with $n_i = 1$, and
$\epsilon_j \geq \mu$ for all $j$ with $n_j = 0$. The distribution of pinning
energies $g(\epsilon)$, now with the intervortex repulsions taken into account
(as opposed to $P$), can be viewed as an {\it interacting} single--particle
density of states and may be obtained from the statistics of the energy levels
$\epsilon_i$; i.e., $g(\epsilon) d\epsilon$ is defined as the number of states
per unit area with energies in the interval $[\epsilon, \epsilon + d\epsilon]$.
Note that in a system of area $L_\perp^2$ this corresponds to a normalization
of the density of states according to
\begin{equation}
	\int_{-\infty}^{+\infty} g(\epsilon) d\epsilon = N_{\rm D} / L_\perp^2
						       = 1 / d^2 \quad .
 \label{dosnor}
\end{equation}
In the following, the terms ``distribution of pinning energies'' and
``(single--particle) density of states'' will be used synonymously. For finite
$\lambda > a_0$, the chemical potential may readily be estimated as
\begin{equation}
	\mu \approx (\lambda / a_0)^2 V(a_0) \approx
      \epsilon_0 f (\lambda / d)^2 \ln \left( f \lambda^2 / d^2 \right) \quad ,
 \label{muesti}
\end{equation}
as each line interacts with approximately $(\lambda / a_0)^2$ other vortices.
Actually, the prefactor of the logarithm in (\ref{muesti}) is independent of
$\lambda$, because $\epsilon_0 \propto 1 / \lambda^2$ \cite{fnote1}. Similarly,
the typical overall width $\gamma$ of the energy level distribution is
\begin{equation}
	\gamma \approx V(a_0) \approx
               \epsilon_0 \ln \left( f \lambda^2 / d^2 \right) \quad ,
 \label{delest}
\end{equation}
and roughly the ``bandwidth'' for the occupied levels will be
$\approx f \gamma$. The relevant energy scales in our problem are thus ordered
according to $\mu \geq f \gamma \gg w$.

Another important quantity is the hopping energy $\Delta_{i \rightarrow j}$
associated with the transfer of a particle from an occupied site $i$ to an
empty site $j$ (conserving the total particle number $N$). The simplest way to
determine this energy cost or gain \cite{cgapth} is to decompose the process
into two steps, namely removing a particle from site $i$ to infinity, thus
gaining the energy $\epsilon_i$, and then taking it from there back into the
system at site $j$, which costs the amount $\epsilon_j - V(r_{ij})$; the
additional contribution $V(r_{ij})$ here stems from the fact that after
removing the particle at site $i$ there were only $N-1$ particles left, while
$\epsilon_j$ was defined for a $N$--particle system, and thus the (fictitious)
interaction with site $i$ has to be accounted for explicitly. Hence one finds
\begin{equation}
	\Delta_{i \rightarrow j} = \epsilon_j - \epsilon_i - V(r_{ij}) \quad ,
 \label{deltai}
\end{equation}
and certainly $\Delta_{i \rightarrow j} > 0$ for all pairs of sites with
$n_i = 1$ and $n_j = 0$ is a necessary condition for the ground state
configuration; in fact, a sufficient condition for a stable ground state may be
stated by demanding that the energies associated with any $m$--particle hops
all be positive. Thus a hierarchical scheme for testing specific configurations
can be constructed, \cite{gapsim} in which one first tests a putative ground
state for stability against one--particle hops, two--particle hops, etc.

With regard to the regime of applicability of the effective Hamiltonian
(\ref{effham}) or (\ref{grcham}), we note once more that for $T > T_1$ entropic
corrections become important, and for $T > T_{\rm dp}$ a flux line can
certainly be no more associated with a single columnar defect, and our
simplified model breaks down. However, the above estimates demonstrate that
thermal fluctuations are in fact essentially negligible up to temperatures very
near the irreversibility line. Similarly, in order that all the flux lines stay
pinned to the columnar defects, the condition $B < B^*(T)$ must be fulfilled,
which provides us with upper limits for the filling $f$, above which the
intervortex repulsions will depin some of the flux lines, and move them to
interstitial sites between the columnar defects (compare Fig.~\ref{phadia}).

Finally, we remark that upon introducing the spin variables
\begin{equation}
	\sigma_i = 2 n_i - 1 \quad ,
 \label{sigmai}
\end{equation}
hence $\sigma_i = +1, -1$ for $n_i = 1, 0$, respectively, the effective
Hamiltonian (\ref{grcham}) is mapped onto a two--dimensional random--site,
random--field antiferromagnetic Ising model with long--range exchange
interactions,
\cite{leesim}
\begin{equation}
	{\tilde H} = {1 \over 2} \sum_{i \not= j} J_{ij} \sigma_i \sigma_j
                               - \sum_i h_i \sigma_i + {\tilde E} \quad ,
 \label{isingm}
\end{equation}
with
\begin{eqnarray}
	J_{ij} &&= {1 \over 4} V(r_{ij}) > 0 \quad ,
 \label{isingj} \\
   h_i &&= {\mu - t_i \over 2} - {1 \over 4} \sum_{j \not= i} V(r_{ij}) \quad ,
 \label{isingh}
\end{eqnarray}
and
\begin{equation}
	{\tilde E} = {1 \over 2} \sum_i \left( t_i - \mu \right) -
  		     {1 \over 8} \sum_{i \not= j} V(r_{ij}) \quad .
 \label{isinge}
\end{equation}
Finding the density of states for this system by any analytical means beyond
mean--field type considerations, \cite{cgapth} or phenomenological scaling
arguments, \cite{ftyexp} constitutes a very difficult task, and has eluded
successful treatment to date. Therefore we have to resort to numerical studies
using a suitable Monte Carlo algorithm; fortunately, the Hamiltonian
(\ref{effham}) is precisely of the form studied in the context of charge
carriers localized at random impurities in doped semiconductors (Coulomb glass
problem), \cite{cgapth,leesim,gapsim} and we may utilize the considerable
experience already gathered in these previous studies.


\subsection{Simulation algorithm}
 \label{simalg}

Our simulation algorithm closely follows the energy--minimization procedure as
described in detail by Efros and Shklovskii in Ref.~\cite{cgapth}, Chap.~14. In
a square of linear size $L_\perp$, a random number generator chooses
$N_{\rm D}$ defect coordinates (by taking $L_\perp = \sqrt{N_{\rm D}}$, the
average defect distance $d$ is held fixed); we have performed simulations for
$N_{\rm D} = 100$, $200$, $400$, and $800$. Then each site is assigned a random
number $t_i$ drawn from the distribution (\ref{sedist}), the width of which we
have chosen as $w / 2 \epsilon_0 = 0.1$, and the interaction potentials
$V(r_{ij})$ are calculated according to (\ref{bespot}) for all pairs $(ij)$.
Following Ref.~\cite{leesim}, we have used periodic boundary conditions, thus
closing the initial square to a torus and thereby reducing boundary effects.
Here, $r_{ij}$ is defined as the minimum value of the distances between site
$i$ and the nine equivalent sites $j$ in the original and repeated ``cells''.
For the interaction range $\lambda$, we have studied the values
$\lambda / d = 0.5, 1, 2, 5$, and $\infty$. In order to handle the latter case
of infinite--range, purely logarithmic interaction carefully, we have applied
the Ewald sum technique, which amounts to including the interaction of each
particle with {\it all} its periodic images; this is achieved by performing the
corresponding series in part in direct, and in part in Fourier space. For
details, we refer to Ref.~\cite{supsol}, App.~A. The data $t_i$ and $V(r_{ij})$
can now be stored for later use.

In the next step, $N = f N_{\rm D}$ ($f < 1$) of these sites $i$ are assigned
the occupation number $n_i=1$. We have employed two very different schemes to
generate this initial configuration: (i) the occupied sites were chosen
randomly, and (ii) a suitable triangular lattice, with a number $N_{\rm lat}$
of sites as close as possible to the correct $N$, was superimposed on the
two--dimensional array of defects, and then deformed until the former lattice
sites matched $N_{\rm lat}$ of the defect positions; the remaining
$N - N_{\rm lat}$ sites were then filled at random. The choices (i) and (ii)
naturally correspond to minimal and maximal initial spatial correlations,
respectively. Using Eq.~(\ref{epsili}), now the site energies $\epsilon_i$ are
calculated. Almost certainly, the initial configuration will correspond to a
highly non--equilibrium situation, in the sense that many of the occupied sites
$i$ will have higher energies $\epsilon_i$ than a large number of empty places
$j$. In order to relax this configuration by successive particle--hole
transitions, the occupied site $p$ of highest energy,
$\epsilon_p = {\rm max}_{\{ n_i=1 \}} \epsilon_i$, and the empty site $q$ of
lowest energy, $\epsilon_q = {\rm min}_{\{ n_i=0 \}} \epsilon_j$ are
determined. If $\epsilon_q < \epsilon_p$, the particle at site $p$ is moved to
site $q$ (the corresponding occupation numbers are interchanged) and the site
energies are recalculated. This procedure is repeated until eventually
$\epsilon_p \leq \epsilon_q$; now the occupied and empty states are separated,
and the chemical potential can be (approximately) obtained from
$\mu = (\epsilon_p + \epsilon_q) / 2$.

However, this intermediate state is still a very poor approximation to the
real ground state, as in general it will be unstable towards single--particle
hops. In order to test this, the hopping energies $\Delta_{i \rightarrow j}$
are calculated for all possible pairs of occupied sites $i$ and empty sites
$j$, using Eq.~(\ref{deltai}). Among those values of
$\Delta_{i \rightarrow j}$, which are negative, the minimum is searched for:
$\Delta_{p \rightarrow q} = {\rm min}_{\{ \Delta_{i \rightarrow j}<0 \}}
\Delta_{i \rightarrow j}$. Next the particle transfer from site $p$ to site $q$
is performed, thus reducing the total energy of the system. Then the new site
energies are calculated, and in general the ``equilibration step'' of the
previous paragraph will have to be repeated. Once more all the
$\Delta_{i \rightarrow j}$ are determined, and the whole procedure is done
again, until finally all the hopping energies acquire positive values. The
ensuing configuration is accepted as a fair approximation to the real ground
state for the particular distribution of defects, filling fraction $f$, and
interaction range $\lambda / d$. The algorithm thus yields the positions of the
$N$ flux lines, the corresponding site energies, and the approximative chemical
potential $\mu$. By repeating this procedure for a number $k$ of different
initial configurations (we took $k = 16000 / N_{\rm D}$), one may then obtain
an ensemble--average of the chemical potential $\langle \mu \rangle$, of the
density of states $\langle g(\epsilon) \rangle$ (from the site energy
statistics), and similarly averaged correlation functions, etc.

Of course, to find the correct ground state, one would in principle have to
test each configuration against any simultaneous $m$--particle hops,
$m = 2,\ldots,\infty$. However, previous investigations have shown that
terminating at $m = 1$ already yields at least qualitatively reliable estimates
for the statistics of the energy level distributions
\cite{cgapth,leesim,gapsim}. In fact, the ``true'' ground state may be very
difficult to reach for the corresponding real system as well, and the static
and dynamic properties may at least at low temperatures be satisfactorily
described by these ``pseudo'' ground states. The quantitative reliability of
the results can furthermore be tested by studying various system sizes, and by
different sampling procedures for the site energies. E.g., one can determine
$\langle g(\epsilon) \rangle$ and $\langle \mu \rangle$ by directly averaging
the energy level statistics of the $k$ different realizations, or by performing
the average on the relative energies ${\tilde \epsilon} = \epsilon - \mu$
calculated in each of the realizations separately. It turns out that these two
versions yield practically identical distributions of pinning energies, the
tiny differences being confined to the immediate vicinity of
$\langle \mu \rangle$, and readily interpreted as finite--size effects (the
second procedure always leads to $g(\epsilon) = 0$ for
$| {\tilde \epsilon} | = | \epsilon - \mu | \leq V(L_\perp)$, because the
lowest possible value of ${\tilde \epsilon}$ is of the order $V(L_\perp)$, see
Ref.~\cite{cgapth}.) One can also check that the two very different initial
conditions described above lead to similar final configurations. In
Sec.~\ref{ivcvrh}, we shall describe how the essentials of the current--voltage
characteristics in the variable--range hopping regime may be obtained from the
thus determined single--particle density of states. (We shall drop the explicit
notation $\langle \ldots \rangle$ for ensemble averages from now on.)

\section{Static Correlations and Transport in the Bose Glass Phase}
 \label{result}


\subsection{Spatial correlations}
 \label{spacor}

We begin the discussion of our results with a description of the spatial
distribution of flux lines among the underlying randomly distributed columnar
pins. Fig.~\ref{xyconf} presents the final configuration in the $xy$ plane
(perpendicular to both the magnetic field and the defect orientation) for
three typical simulation runs, performed on the identical defect configuration
($N_{\rm D} = 400$), with infinite--range logarithmic interaction
($\lambda \rightarrow \infty$), and using site energies drawn from the flat
distribution (\ref{sedist}) with $w / 2 \epsilon_0 = 0.1$ (compare the
estimates in Sec.~\ref{modeqs}), but with different filling fractions
$f = 0.1$, $f = 0.2$, and $f = 0.4$, respectively, in each case starting from a
random initial distribution of vortices (the final states stemming from the
other extreme of possible initial conditions, namely a distorted triangular
lattice, are not identical but look qualitatively very similar, see below). In
contrast to the Poisson distribution of defects (open circles), showing the
characteristic voids as well as clusterings, the flux lines (full circles) form
a highly correlated structure, trying to keep as distant as possible from each
other to minimize the interaction energy. For low fillings
[Figs.~\ref{xyconf}(a) and (b)] the ensuing spatial distribution of flux lines
resembles a distorted triangular lattice. However, for higher filling
fractions, $f \geq 0.4$ say, the vortices are increasingly forced to accomodate
the underlying random pin distribution. As can be seen in Fig.~\ref{xyconf}(c),
they then tend to occupy the energetically favorable sites encircling spatial
voids in the defect distribution. Upon increasing $f$ even more, beyond
$f \geq 0.7$ say, (which may be achieved by applying higher magnetic fields
$B$) the spatial correlations disappear on longer length scales, and the only
remaining effect of the intervortex repulsion is to prevent simultaneous
occupation of nearest--neighbor sites. One has to keep in mind, however, that
in this regime our basic assumption that each flux line be attached to a single
defect, which requires $B < B^*(T)$ [Eq.~(\ref{crossb})], is beginning to break
down. As has been explained in Sec.~\ref{modeqs}, for $B > B^*(T)$ interactions
will become important in determining the vortex localization length, and
eventually force some of the flux lines to leave the columnar defects and move
to interstitial sites. Most of our numerical work is restricted to $f \leq 0.4$
for this reason. We note that all of these features do not very strongly depend
on the actual value of the parameter $\lambda / d$, as long as
$\lambda / d \geq 1$, which ensures that the typical interaction energies
exceed the fluctuations in pinning potential depths $w$.

More quantitative conclusions can be drawn from the calculation of the static
structure factor $S({\bf q})$,
\begin{equation}
	S({\bf q}) =
              {1 \over N} \sum_{i,j = 1}^N e^{i {\bf q}({\bf r}_i-{\bf r}_j)} =
	      {1 \over N} | n({\bf q}) |^2 \quad ,
 \label{strfac}
\end{equation}
which is readily obtained from the two--dimensional Fourier transform
$n({\bf q}) = \sum_i e^{i {\bf q}{\bf r}_i}$ of the vortex density
$n({\bf r}) = \sum_i \delta({\bf r} - {\bf r}_i)$; more precisely, we shall
consider the average of (\ref{strfac}) over the directions in momentum space,
\begin{equation}
	S(q) = \int d \Omega_{\bf q} \, S({\bf q})
             = \int_0^{2 \pi} d \phi \, S(q,\phi) \quad .
 \label{sqaver}
\end{equation}
In Fig.~\ref{sqconf}, the calculated $S(q)$ is depicted for both the initial
and final vortex configurations, for filling fractions $f = 0.1$ (a), and
$f = 0.2$ (b). These curves were obtained from averaging over $k = 40$
different runs with $N_{\rm D} = 400$, $w / 2 \epsilon_0 = 0.1$, and
$\lambda / d = 5$. Again, the corresponding pictures for any
$\lambda / d \geq 1$ (keeping $w$ fixed) look very much alike. Note that the
peak in the dashed curve in Fig.~\ref{sqconf}(a) at
$q a_0 \approx 4 \pi / \sqrt{3}$, i.e., at the Bragg peak corresponding to the
triangular lattice, is shifted to lower values and diminished in height in
Fig.~\ref{sqconf}(b); for higher fillings the underlying random pin
distribution enforces an average separation of vortices larger than the ideal
hexagonal close--packed geometry would permit. For similar reasons, the peak of
$S(q)$ for the equilibrated configurations (corresponding to occupied sites in
Fig.~\ref{xyconf}) is displaced to lower values with respect to the ideal
triangular lattice, namely to $q a_0 \approx 0.86 (4 \pi / \sqrt{3})$, and also
broadened considerably due to the appearance of a large number of topological
defects (see Fig.~\ref{vrlami} below). Interestingly, for $f = 0.1$ there is a
double--peak structure. The overall appearance, width and peak height of the
dotted and solid curves in Fig.~\ref{sqconf} are very similar, suggesting the
relative unimportance of the specific choice of initial conditions. In addition
to the suppression of fluctuations at low $q a_0 \leq 4$ and the marked peak in
$S(q)$ at (roughly) $q a_0 \approx 2 \pi$, the shallow dip in the interval
$9 \leq q a_0 \leq 12$ should be noted. For these spatial correlation effects,
the most sensitive parameter turns out to be the filling fraction $f$, as
illustrated in Fig.~\ref{sqfill} for $\lambda / d = 2$. While for $f = 0.1$ and
$f = 0.2$ the structure factor is practically indistinguishable from the
previous pictures for $\lambda / d = 5$, already for $f = 0.4$ the peak height
becomes considerably reduced, and the shallow dip has disappeared. For
$f = 0.8$ the peak has vanished, and even the correlation gap at low $q$ is
halfway closed, thus rendering $S(q)$ more and more uniform as for a Poisson
distribution (cf. the long--dashed curves in Fig.~\ref{sqconf}).

Returning to the configuration plots in direct space (Fig.~\ref{xyconf}), we
can analyze the final structures more thoroughly by performing a Delaunay
triangulation, thereby exposing the topological defects (with respect to an
ideal six--fold coordinated triangular lattice), simply by determining the
ensuing coordination number $N_c$ at each vortex site. The results of this
procedure for the configurations of Figs.~\ref{xyconf}(a)--(c) are shown in
Fig.~\ref{vrlami}. Clearly the resulting vortex configuration can only be
viewed as a highly distorted ``triangular lattice'', with its orientational
correlations being strongly diminished.

In addition to these gross interaction effects, there emerge subtle
correlations in the single--particle energies, namely the spatial clustering of
those sites whose energies $\epsilon_i$ differ very little from the chemical
potential $\mu$. This remarkable effect has previously been observed by Davies,
Lee, and Rice for the Coulomb glass case \cite{leesim}. We illustrate this
property for the distribution in Fig.~\ref{xyconf}(c), i.e.: $N_{\rm D} = 400$,
$w / 2 \epsilon_0 = 0.01$, $\lambda \rightarrow \infty$, and $f = 0.4$. In
Fig.~\ref{cllami}, along with the entire configuration, the spatial
distribution of those filled and empty sites with energies in the range
$| \epsilon_i - \mu | \leq \delta$ are shown, with
$\delta / 2 \epsilon_0 = 0.6$ (a), and $\delta / 2 \epsilon_0 = 1.2$ (b),
respectively [compare Fig.~\ref{dslami}(b) below]. The occupied sites with
energies close to $\mu$ are strikingly clustered in space, and the
corresponding low--energy empty sites seem to occupy the complementary region.
The interactions cause long--range spatial fluctuations in the pinning
energies, which are apparently yet not understood in detail. Of course, the
very fact that a certain occupied site attains a high site energy must mean
that there are other occupied sites nearby. It is less clear, however, why the
energies of nearby occupied sites for $\delta \ll \mu$ should be so similar, or
why the states in the low--energy part of the empty pseudo--band should be
similarly correlated in space. Note that the Bose glass is not symmetric with
respect to exchange of particles and holes, since the empty sites do not
interact. Thus, defining ${\tilde \epsilon}_i = \epsilon_i - \mu$, those states
$i,j$ with both $| {\tilde \epsilon}_i |$ and
$| {\tilde \epsilon}_j | \leq \delta$, are situated nearby when
${\tilde \epsilon}_i{\tilde \epsilon}_j > 0$, and are widely separated in space
when ${\tilde \epsilon}_i{\tilde \epsilon}_j < 0$. Upon increasing the
threshold $\delta$, both the occupied and the empty
low--$| {\tilde \epsilon}_i |$ clusters appear to percolate through the system,
finally becoming increasingly mixed for $\delta \geq \mu$, and the complete
configuration is reached.

Recently, the positions of both columnar defects and flux lines have been
determined simultaneously in experiments on ${\rm NbSe_2}$ \cite{stmexp} and
BSCCO samples \cite{decexp}, using scanning tunneling microscopy and a
combination of chemical etching and magnetic decoration techniques. Indeed, the
manifest spatial correlations in Fig.~4 of Ref.~\cite{decexp} strongly suggest
the relevance of vortex interactions. In a recent note, \cite{letter} we used
such experimental data on the distribution of pins and vortices to infer the
density of states and the ensuing transport characteristics, which are thus
subject to direct verification by measurements (the relevant parameters in that
case turned out to be $f \approx 0.24$, $\lambda / d \approx 0.96$, and
$w \approx 0.1 \epsilon_0$). This possibility of quantitative comparison of the
spatial configurations and correlations as obtained from simulations with those
measured in actual experiments opens a fascinating new field for detailed
studies, which have been far from feasible in the otherwise closely related
disordered semiconductor systems \cite{cgapth}. Even the striking clustering in
Fig.~\ref{cllami} may possibly be observed experimentally, if the current
techniques of manipulating individual vortices with the tip of a magnetic force
microscope \cite{vorman} are further refined, which might eventually permit the
measurements of individual pinning energies $\epsilon_i$.


\subsection{Density of states and Coulomb gap}
 \label{dosgap}

Before we proceed to present our simulation results for the single--particle
density of states (i.e.: the distribution of vortex pinning energies, taking
their interactions into account), we briefly summarize Efros and Shklovskii's
qualitative arguments that long--range repulsive forces lead to a soft gap in
$g(\epsilon)$ \cite{cgapth}. Let us consider particles in $D$ dimensions, which
are localized on randomly distributed sites ${\bf r}_i$, and interact via an
arbitrary potential
\begin{equation}
	V(r_{ij}) = 2 \epsilon_0 K(r_{ij}/\lambda) \quad ,
 \label{lrptl}
\end{equation}
where, quite generally, $\epsilon_0$ sets the energy scale, $\lambda$ defines a
characteristic length, and $K(x)$ is a positive, monotonically decreasing
function approaching zero for $x \rightarrow \infty$. Its inverse therefore
exists, and shall be denoted by ${\tilde K}(x) = K^{-1}(x)$.

The simplest argument to determine the interacting density of states now is as
follows (see Chap.~10 of Ref.~\cite{cgapth}). Consider site energies
$\epsilon_i$, $\epsilon_j$ located in an interval of width ${\tilde \epsilon}$
around the chemical potential $\mu$, but situated on opposite sides of $\mu$,
e.g. with $n_i = 1$ and $n_j = 0$. Then in the ground state a particle transfer
from site $i$ to $j$ necessarily costs the positive energy
$\Delta_{i \rightarrow j} = \epsilon_j - \epsilon_i - V(r_{ij}) > 0$
[Eq.~(\ref{deltai})], which implies $V(r_{ij}) < \epsilon_j - \epsilon_i \leq
{\tilde \epsilon}$. Therefore, hops typically have to occur over a distance
\begin{equation}
	r_{ij}({\tilde \epsilon}) \geq R({\tilde \epsilon}) \approx
		\lambda {\tilde K}({\tilde \epsilon} / 2 \epsilon_0) \quad .
 \label{typhop}
\end{equation}
The concentration of relevant, i.e.: energetically available, impurities hence
is $n({\tilde \epsilon}) \propto R({\tilde \epsilon})^{-D}$. Thus for small
${\tilde \epsilon}$ the density of states $g({\tilde \epsilon}) \propto
d n({\tilde \epsilon}) / d {\tilde \epsilon}$ becomes
\begin{equation}
      g({\tilde \epsilon}) = c_D {1 \over 2 \epsilon_0 \lambda^D}
			  {| {\tilde K}'(x) | \over {\tilde K}(x)^{D+1}}
                     \Bigg \vert_{x = {\tilde \epsilon} / 2 \epsilon_0} \quad ,
 \label{gengap}
\end{equation}
where $c_D$ is a proportionality factor of order $O(1)$.

We can now specialize to the true long--range potentials $K(x) = x^{-\sigma}$
with $0 < \sigma < D$ and $K(x) = - \ln x$ (which formally results from the
previous case in the limit $\sigma \rightarrow 0$). In these situations, the
prefactors $c_D$ can actually be determined using the somewhat refined
self--consistency argument by Efros \cite{sccgap}, which yields
$c_D(\sigma) = 2 D (1 - \sigma / D) \Gamma(D/2+1) \pi^{-D/2}$.
Eq.~(\ref{gengap}) then gives for the inverse--power interactions
\begin{equation}
	g({\tilde \epsilon}) = {D \Gamma(D/2+1) \over
				\sigma \pi^{D/2} \lambda^D \epsilon_0}
 \left( {{\tilde \epsilon} \over 2 \epsilon_0} \right)^{D / \sigma - 1} \quad ,
 \label{powgap}
\end{equation}
i.e.: the density of states near the chemical potential vanishes as a power law
[cf. Eq.~(\ref{gapexp})] with gap exponent $s_0(D) = D / \sigma - 1$; e.g.: for
the Coulomb potential ($\sigma = 1$) in $D = 2$ and $D = 3$ dimensions one has
$s_0(2) = 1$ and $s_0(3) = 2$, respectively. For the purely logarithmic
interaction, one similarly finds
\begin{equation}
  g({\tilde \epsilon}) = {D \Gamma(D/2+1) \over \pi^{D/2} \lambda^D \epsilon_0}
	\exp \left( {D {\tilde \epsilon} \over 2 \epsilon_0} \right) \quad ;
 \label{loggap}
\end{equation}
i.e.: although $g(\epsilon)$ does not vanish for $\epsilon = \mu$, there is an
exponential dip near the chemical potential. Naturally, there exists a certain
upper cut--off for large ${\tilde \epsilon}$, beyond which the density of
states assumes its usual shape.

It is important to note that Eqs.~(\ref{gengap})--(\ref{loggap}), need not be
quantitatively correct. Correlations in the distribution of site energies have
been neglected, and the long--range fluctuations embodied in the remarkable
clustering described at the end of the previous subsection (Fig.~\ref{cllami})
are not accounted for. Also, only single--particle transfers were considered in
the above discussion, and multi--particle hops could possibly modify the
results, especially in the case of purely logarithmic interactions
\cite{ftyexp}. In fact, Monte Carlo simulations using variants of the
above--described algorithm \cite{leesim,gapsim} yield gap exponents somewhat
larger than the mean--field estimates $s_0(D)$. E.g., the most detailed study
of the Coulomb glass performed by M\"obius, Richter, and Drittler found
$s(2) = 1.2 \pm 0.1$ and $s(3) = 2.6 \pm 0.2$ in $D = 2$ and $D = 3$
dimensions, respectively, for the best fits to their results on very large
systems (up to 40 000 and 125 000 sites, respectively). These authors,
moreover, were ultimately unable to decide whether the density of states near
the chemical potential is correctly described by a power law at all. Although
Eq.~(\ref{loggap}) for logarithmic repulsions may have to be altered as well, a
power--law fit in the energy--range accessible to our simulations appears not
unreasonable.

In Fig.~\ref{dslami}, we present some of the simulation results for the density
of states of our effectively two--dimensional system with the interaction
(\ref{bespot}), in the limit $\lambda \rightarrow \infty$, i.e.: for a purely
logarithmic potential. They were obtained by sampling the site energies of
$k = 40$ runs each with either random initial conditions (open circles) or
starting out with a distorted triangular lattice (filled triangles), using a
system size of $N_{\rm D} = 400$ defect sites and a random site energy
distribution of width $w / 2 \epsilon_0 = 0.1$. In this case of infinite--range
interactions, we have used the Ewald summation procedure in order to obtain
reliable results, and before taking the limit $\lambda \rightarrow \infty$ a
term $\propto \lambda^2$ has been subtracted from the site energies,
\cite{fnote2}
\begin{equation}
	\epsilon_i' = \epsilon_i - 2 \pi f (\lambda / d)^2 \quad .
 \label{epsshi}
\end{equation}
The specific choice of the initial conditions for the Monte Carlo algorithm
has hardly any influence. Although there is a general tendency for the random
initial configurations to yield lower total energies on average, this slight
difference lies well in the range of the actual fluctuations in $\mu$ over the
$k = 40$ runs with different defect distributions. At any rate, the emerging
soft ``Coulomb'' gap near the chemical potential is very distinct and
surprisingly broad for both cases $f = 0.2$ [Fig.~\ref{dslami}(a)] and
$f = 0.4$ [Fig.~\ref{dslami}(b)]. The overall width of the site--energy
distribution is determined by the interactions, and slightly increases with
$f = N / N_{\rm D} = B / B_\phi$ [Eq.~(\ref{delest})]. As the filling fraction
grows, however, the width of the Coulomb gap itself (at half maximum, say) does
not change. Its relative width thus decreases sowewhat with larger fillings, as
it ``sweeps through'' the density of states along with the chemical potential.
More important, the effective gap exponent apparently depends monotonically on
$f$, attaining its largest values for small fillings.

This becomes more clearly visible in the double--logarithmic plots of
Fig.~\ref{ldlami}, where both the filled (full symbols) and empty (open
symbols) ``subbands'' have been folded onto the same side. From the single
decade of data showing the increase of the density of states with growing
distance from $\mu$, we can infer the effective gap exponents
$s_{\rm eff} \approx 2.9$ for $f = 0.1$ [Fig.~\ref{ldlami}(a)], and
$s_{\rm eff} \approx 2.2$ for $f = 0.4$ [Fig.~\ref{ldlami}(b)]. The apparent
dependence on $f$ arises because upon increasing the filling the flux lines
have to accomodate more and more with the underlying random pin distribution,
whereby correlations are destroyed. Of course, we cannot be sure that we have
reached an asymptotic $f$--independent exponent with our simulations. It
nevertheless seems clear that the mean--field formula (\ref{loggap}) is
inadequate, and that a power law of the form (\ref{gapexp}) provides a better
description of the soft Coulomb gap, with a maximum gap exponent
\begin{equation}
	s_{\rm eff}^{\rm max}(f \ll 1) \approx 3 \quad .
 \label{mxgexp}
\end{equation}
As is implied by our notation, we do not consider these gap exponents as
universal quantities; rather, they should be viewed as {\it effective}
exponents describing the shape of the Coulomb gap in some specific energy
range.

The details of the asymptotic analytical form of $g(\epsilon)$ in the limit
$\epsilon \rightarrow \mu$ are not our main concern. We are most interested in
the striking shape of the density of states, displaying a very marked and wide
correlation--induced ``Coulomb'' gap, because this has a dramatic
{\it qualitative} effect on the current--voltage characteristics (see
Sec.~\ref{ivcvrh}). One of the most important features of the pseudo--gap
of Fig.~\ref{dslami} in fact rests in its remarkable persistence as the
interaction range, set by the London penetration depth $\lambda$, becomes
finite. Figs.~\ref{dslam5}--\ref{dslam1} depict a selection of results for the
single--particle density of states for decreasing $\lambda$, with various
fillings $f$ (and the other parameters chosen as before). As can be seen in
Fig.~\ref{dslam5}, the situation with $\lambda / d = 5$ is practically
indistinguishable from the previous case of infinite interaction range, in
overall appearance, width and general distribution of pinning energies [cf.
Fig.~\ref{dslami}(b) and Fig.~\ref{dslam5}(b) for identical $f = 0.4$]. The
overall curves are simply shifted along the energy axis according to
Eq.~(\ref{epsshi}) (see also Fig.~\ref{mulam5} below). In Fig.~\ref{dslam5} the
more extreme cases with (a) $f = 0.1$, with a very broad and flat pseudogap,
and (c) $f = 0.8$ are shown as well. In the latter case, $g(\mu)$ is manifestly
non--zero. This is due in part to the finiteness of $\lambda$, thereby
subjecting the system more and more to the underlying pinning potential
randomness. In addition, the fluctuations in the chemical potential grow
considerably as $f$ is increased, thus obscuring the vanishing $g(\mu)$ by
sampling over the energies $\epsilon_i$ \cite{fnote3}. Nevertheless, it is
clear that the Coulomb gap closes: its width becomes narrower and
$g(\mu) \propto \lambda^{-2}$ grows, as may be seen by comparing
Figs.~\ref{dslam5}(a) and \ref{dslam2}(a) for $f = 0.1$, Figs.~\ref{dslami}(a)
and \ref{dslam1}(a) for $f = 0.2$, Figs.~\ref{dslami}(b), \ref{dslam5}(b), and
\ref{dslam2}(b) for $f = 0.4$, and Figs.~\ref{dslam5}(c) and \ref{dslam1}(b)
for $f = 0.8$. Even for $\lambda \approx d \leq a_0$ there remains a
considerable suppression of $g(\mu)$ itself, accompanied by a strong depletion
of the density of states up to $| {\tilde \epsilon} | / 2 \epsilon_0
\approx 0.2$. In an intermediate range of ${\tilde \epsilon}$, one may still
use Eq.~(\ref{gapexp}), and the effective gap exponent for $\lambda / d = 1$
and $f = 0.1$, for example, is about $s_{\rm eff} \approx 1.2$. Notice that in
accord with our estimate (\ref{muesti}), the chemical potential roughly scales
as $\propto f (\lambda / d)^2$ as filling and interaction range are varied.


\subsection{Finite size effects}
 \label{finsiz}

We now discuss finite size effects, i.e.: the dependence on the number of
defect sites $N_{\rm D}$ used in the Monte Carlo simulations. We have performed
extensive studies for the cases $N_{\rm D} = 100$, $200$, $400$ (for which all
the previous plots are depicted), and $800$, averaging over
$k = 16000 / N_{\rm D}$ runs with different initial defect configurations (thus
supplying an equal number of data points). The total width and shape of the
density of states, as well as the appearance of the Coulomb gap, remained
practically identical upon changing $N_{\rm D}$ in the above range. The sole
effect is a chemical potential shift, or equivalently, the total energy per
particle, for the situations with long--range interactions. For
$\lambda / d = 1$, $\mu(N_{\rm D})$ is independent of $N_{\rm D}$ for
$N_{\rm D} \geq 200$. In Fig.~\ref{mulam5}, the mean values for the obtained
chemical potentials are plotted as a function of $1 / N_{\rm D}$ for the runs
with $\lambda / d = 5$ and several filling fractions $f \leq 0.5$. As is to be
expected, upon increasing the number of sites and consequently interacting
particles, for this very long--range repulsive forces $\mu(N_{\rm D})$ grows,
but the data approach finite values in the limit
$N_{\rm D} \rightarrow \infty$, close to the open symbols on the
$1 / N_{\rm D} = 0$ axis. These latter points were obtained by shifting the
results for $\mu(N_{\rm D} = 400)$ of the $\lambda \rightarrow \infty$
simulations (obtained via the Ewald summation procedure) according to
Eq.~(\ref{epsshi}), with $\lambda / d = 5$. Thus the size dependence of the
chemical potential is easily understood; and there seem to be few other
finite--size artifacts except that we cannot sample energies closer to $\mu$
than ${\tilde \epsilon} \approx V(L_\perp) \approx 2 {\tilde \epsilon}
K_0(L_\perp / \lambda)$ (see Sec.~\ref{simalg}). Note that $\mu \propto f$ in
Fig.~\ref{mulam5}, which is equivalent to $B \approx H$ in the flux--line
language.


\subsection{Current--voltage characteristics in the variable--range hopping
            regime}
 \label{ivcvrh}

We now determine how the depletion in $g(\epsilon)$ near the chemical potential
affects vortex transport properties. An in--plane current
${\bf J} \perp {\bf B}$ induces a Lorentz force per unit length
${\bf f}_{\rm L}$ perpendicular to ${\bf J}$, acting on all the flux lines:
\begin{equation}
    {\bf f}_{\rm L} = {\phi_0 \over c} \, {\hat {\bf z}} \times {\bf J} \quad .
 \label{lorfor}
\end{equation}
Accordingly, we add the term
\begin{equation}
	\delta F_N[\{ {\bf r}_i \}] =
		- \int_0^L dz \sum_{i=1}^N {\bf f}_{\rm L} {\bf r}_i(z)
 \label{lorffe}
\end{equation}
to our model free energy (\ref{modelf}). As stated above, in the boson picture
this additional term represents an electric field
${\bf E} = {\hat {\bf z}} \times {\bf J} / c$ acting on the particles carrying
charge $\phi_0$ (see Table \ref{bosmap}). This fictitious quantity ${\bf E}$
should not be confused with the true electric field ${\cal E}(J)$ (the voltage
drop induced by the current through the movement of flux lines). In the spirit
of the thermally assisted flux--flow (TAFF) model of vortex transport,
\cite{taffmd} the superconducting resistivity (i.e.: the conductivity in the
boson representation) $\rho \equiv {\cal E} / J$ may be written as
\begin{equation}
	\rho \approx \rho_0 \exp \left[ - U_{\rm B}(J) / T \right] \quad ,
 \label{taffre}
\end{equation}
where $\rho_0$ is a characteristic flux--flow resistivity, and $U_{\rm B}$
represents an effective barrier height. Unlike the original TAFF models,
$U_{\rm B}(J)$ actually diverges as $J \rightarrow 0$ in the Bose glass phase
[see Eq.~(\ref{ivchar})].

Consider first the regime where the motion of a single vortex is unaffected by
the other flux lines in the sample. We shall see that this is true only in an
intermediate current regime $J_1 < J < J_c$ \cite{nelvin}. Driven by the
external current $J$, a flux line will start to leave its columnar pin by
detaching a segment of length $z$ into the defect--free region, thereby forming
a half--loop of transverse size $r$. The free energy price for this process
consists of the elastic energy $\approx {\tilde \epsilon}_1 r^2 / z$, and the
lost pinning energy $\approx U_0 z$, as the first vortex to move will be the
most weakly bound one. Note that we have to re--introduce the average pinning
potential $U_0 = \langle U_k \rangle$ here. Upon taking into account the work
done by the Lorentz force, the free energy of the loop is approximately
\cite{nelvin}
\begin{equation}
	\delta F_1(r,z) \approx {\tilde \epsilon}_1 r^2 / z + U_0 z
			- f_{\rm L} r z \quad .
 \label{feloop}
\end{equation}
Optimizing $\delta F_1(r,z)$ for $f_{\rm L} = 0$ first, we see that for the
saddle--point configuration
\begin{equation}
	z^* \approx r^* \sqrt{{\tilde \epsilon}_1 / U_0} \quad ,
 \label{loopfm}
\end{equation}
while for finite currents
\begin{equation}
	r^* \approx c U_0 / \phi_0 J \quad ,
 \label{loopsz}
\end{equation}
which yields the saddle--point free energy
\begin{equation}
   \delta F_1^* \approx c {\tilde \epsilon}_1^{1/2} U_0^{3/2} / \phi_0 J
	\quad ,
 \label{loopfe}
\end{equation}
which we may identify as the energy barrier in Eq.~(\ref{taffre}) for the
half--loop nucleation process \cite{nelvin}. For a sufficiently low current
$J_1$, the half--loop will typically extend to the nearest--neighbor pin,
namely when $r^* \approx d$, and hence
\begin{equation}
	J_1 = c U_0 / \phi_0 d \quad .
 \label{loopcr}
\end{equation}
The flux line will then form a double--kink of width
\begin{equation}
	w_{\rm K} = d \sqrt{{\tilde \epsilon}_1 / U_0} \quad ,
 \label{kinkwd}
\end{equation}
which costs a free energy $2 E_{\rm K}$, where
\begin{equation}
	E_{\rm K} = d \sqrt{{\tilde \epsilon}_1 U_0} \quad .
 \label{kinken}
\end{equation}
These expressions allow us to cast the current--voltage characteristics for
$J_1 < J < J_c$ into the form \cite{nelvin}
\begin{equation}
	{\cal E} \approx
	\rho_0 J \exp \left[ - (E_{\rm K} / T) (J_1 / J) \right] \quad .
 \label{ivloop}
\end{equation}
In the half--loop regime, the intervortex repulsions thus do not affect the
transport exponent $p_{\rm eff} = 1$, cf. Eq.~(\ref{ivchar}). In recent
experiments on BSCCO samples, exponents $p_{\rm eff} \approx 1$ have been found
for intermediate currents, supporting the half--loop nucleation picture
\cite{trexp2,trexp3}.

For $J < J_1$ we thus have to take the configurational limitations imposed by
the other vortices into account. The most important thermally activated
excitation will now be a double superkink (Fig.~\ref{vrkink}): the flux line
will send out a tongue--like pair of kinks to another possibly very distant
defect, which has a favorable pinning energy, thus optimizing the tunneling
probability. This is the flux line analogue of variable--range charge transport
in disordered semiconductors \cite{cgapth}. The cost in free energy for such a
configuration of transverse size $R$ and extension $Z$ along the
magnetic--field direction will consist of two terms: (i) the double--superkink
energy $2 E_{\rm K}(R) = 2 E_{\rm K}(d) R / d$ stemming from the elastic term,
and (ii) the difference in pinning energies of the highest--energy occupied
site, $\epsilon_i \approx \mu$, and the empty site at distance $R$ with
$\epsilon_j = \mu + \Delta(R)$. Thus the free energy difference with respect to
the situation without kinks and external current is
\begin{equation}
    \delta F_{\rm SK} \approx 2 E_{\rm K} R / d + Z \Delta(R) - f_L R Z \quad ,
 \label{fesknk}
\end{equation}
which generalizes Eq.~(4.13) of Ref.~\cite{nelvin} for the case of
non--interacting flux lines. The concentration available states as a function
of $R$ with $D$ dimensions transverse to ${\bf B}$ (here $D = 2$) on the one
hand equals $d^D \int_\mu^{\mu + \Delta(R)} g(\epsilon) d \epsilon$, and on the
other hand is simply given by $\approx (d / R)^D$; thus $\Delta(R)$ is to be
determined from the equation
\begin{equation}
	\int_\mu^{\mu + \Delta(R)} g(\epsilon) d \epsilon = R^{-D} \quad .
 \label{deltar}
\end{equation}

Optimizing now first for vanishing current $J = 0$ gives the longitudinal
extent $Z^*$ of the superkink as a function of its transverse size $R^*$,
\begin{equation}
	Z^* \approx
	- {2 E_{\rm K} / d \over (\partial \Delta / \partial R)_{R^*}} \quad .
 \label{longsz}
\end{equation}
Upon balancing the last term in Eq.~(\ref{fesknk}) against the optimized sum of
the first two, one arrives at
\begin{equation}
	J \phi_0 / c \approx \Delta(R^*) / R^* \quad ,
 \label{transz}
\end{equation}
which through inversion yields a typical hopping range $R^*(J)$. Inserting back
into (\ref{fesknk}) finally yields the result for the optimized free energy
barrier for a jump,
\begin{equation}
	\delta F_{\rm SK}^*(J) \approx (2 E_{\rm K} / d) R^*(J) \quad ,
 \label{deltaf}
\end{equation}
which we identify with the current--dependent activation energy in
Eq.~(\ref{taffre})
\begin{equation}
	{\cal E} \approx
	\rho_0 J \exp \left[ - (2 E_{\rm K} / T d) R^*(J) \right] \quad .
 \label{ivsknk}
\end{equation}

Consider now a power--law form for the distribution of pinning energies,
$g(\epsilon) = \kappa | \epsilon - \mu |^s$. Then from Eq.~(\ref{deltar})
\begin{equation}
   \Delta(R) = \left( {s+1 \over \kappa} \right)^{1/(s+1)} R^{-D/(s+1)} \quad ,
 \label{pldelr}
\end{equation}
and (\ref{transz}) yields
\begin{equation}
	R^*(J) = \left( {s+1 \over \kappa} \right)^{1/(D+s+1)}
		 \left( {c \over \phi_0 J} \right)^p \quad ,
 \label{pltrsz}
\end{equation}
where
\begin{equation}
	p = {s + 1 \over D + s + 1} \quad .
 \label{plivex}
\end{equation}
This immediately leads to
\begin{equation}
	\delta F_{\rm SK}^*(J) = 2 E_{\rm K} (J_0 / J)^p \quad ,
 \label{pldelf}
\end{equation}
which is of the form (\ref{ivchar}) with the transport exponent (\ref{plivex})
and the current scale
\begin{equation}
	J_0 \approx c / \phi_0 \kappa^{1/(s+1)} d^{1/p} \quad .
 \label{plcrsc}
\end{equation}
For a constant density of states, $s = 0$ and $\kappa = g(\mu)$, one recovers
the non--interacting results, namely the Mott variable--range hopping exponent
in $D$ dimensions $p_0 = 1 / (D + 1)$, and expression (\ref{curscl}) for $J_0$
\cite{nelvin}. Using the mean--field result for the gap exponent,
$s_0 = D / \sigma - 1$ [Eq.~(\ref{powgap})], Eq.~(\ref{plivex}) gives
$p = 1 / (1 + \sigma)$, i.e.: $p \rightarrow 1$ for purely logarithmic
interactions. In the light of our results in Sec.~\ref{dosgap}, however, we
have to expect that $p_{\rm eff}$ will actually be smaller; namely with
Eqs.~(\ref{mxgexp}) and (\ref{plivex}) we find
\begin{equation}
	p_{\rm eff}^{\rm max} \approx 2 / 3
 \label{maxexp}
\end{equation}
in two dimensions, for $\lambda \rightarrow \infty$ and small fillings.

The numerical result (\ref{maxexp}) for the variable--range hopping transport
exponent of logarithmically interacting particles in two dimensions is
consistent with the analysis by Fisher, Tokuyasu, and Young, \cite{ftyexp}
although the underlying theoretical arguments appear to be different. One might
also question whether the ``gauge glass'' model studied in Ref.~\cite{ftyexp},
with its quenched random vector potential, correctly captures the physics of a
finite density of, say, positive vortices subject to a random pinning
potential.

Using the general relations (\ref{deltar})--(\ref{ivsknk}) and the Monte Carlo
results of Sec.~\ref{dosgap} for the single--particle density of states, we can
numerically evaluate the current--voltage characteristics for any form of
$g(\epsilon)$. Results derived from the distributions of interacting pinning
energies in Figs.~\ref{dslami}--\ref{dslam1} are depicted in Fig.~\ref{ivfill}
as double--logarithmic plots of the effective hop size
$R^*(J) \propto \delta F_{\rm SK}^*(J) = U_{\rm B}(J)$ as function of the
current $J$; according to Eq.~(\ref{ivsknk}), the slope immediately yields the
effective transport exponent $p_{\rm eff}$ [cf. Eq.~(\ref{pltrsz})]. Each plot
corresponds to a certain filling fraction $f$, and summarizes the results for
various values of the parameter $\lambda / d$. In Fig.~\ref{ivfill}(a) for
$f = 0.1$ the curve corresponding to long--range repulsions ($\lambda / d = 5$,
which turns out to be practically indistinguishable from the case
$\lambda \rightarrow \infty$) yields $p_{\rm eff} \approx 0.70 \pm 0.05$, in
accord with (\ref{maxexp}). Upon decreasing the interaction range,
$p_{\rm eff}$ becomes smaller, $p_{\rm eff} \approx 0.55 \pm 0.05$ for
$\lambda / d = 2$ and $p_{\rm eff} \approx 0.50 \pm 0.05$ for
$\lambda / d = 1$, and finally reaches the non--interacting exponent
$p_0 = 1/3$ for $\lambda \rightarrow 0$. Note that there appears as well a
considerable shift in the prefactor for the power law which also enhances the
effectiveness of the vortex pinning, when the repulsive interactions are turned
on. Long--range repulsive interactions thus reduce vortex voltages
$\propto \exp[- 2 E_{\rm K} R^*(J) / d T]$ by several orders of magnitudes with
respect to the non--interacting situation. E.g., for $\log j \approx 1.5$ the
exponent in (\ref{ivsknk}) is about ten times smaller for $\lambda / d = 5$ as
compared to $\lambda \rightarrow 0$, and hence the resistivity is reduced by a
factor of $\approx 10^{-5}$ [note that $\mu$ as function of $\lambda$ does
actually not vary very much, as $\epsilon_0 \propto \lambda^{-2}$, see
Eq.~(\ref{muesti})].

For higher fillings $f$, this collective pinning mechanism becomes relatively
weaker, as the flux lines increasingly have to accomodate the underlying random
pin distribution, and correlations are destroyed (see the discussion in
Sec.~\ref{dosgap}). This can be seen in Fig.~\ref{ivfill}(b) and (c); for
$f = 0.8$ all the curves with different finite $\lambda$ are essentially
parallel to the the one for $\lambda \rightarrow 0$, i.e.:
$p_{\rm eff} \approx 1/3$. However, there is still a considerable offset due to
the change in the prefactor of the power law, and a reduction of the
exponential factor in (\ref{ivsknk}) by about three, and hence $\rho$ still
about twenty times smaller than for the case without interactions. We remark
that of course the data for the very lowest currents in Fig.~\ref{ivfill} are
less reliable, for the finite size of our system becomes important when
$R^*(J) \approx L_\perp$, which for $N_{\rm D} = 400$ happens for
$\log [ R^*(J) / d ] \approx 1.3$.

In at least one recent experiment on BSCCO, reported by Konczykowski,
Chikumoto, Vinokur, and Feigel'man, \cite{trexp2} an effective transport
exponent $p_{\rm eff}$ in the variable--range hopping regime clearly different
from the non--interacting $p_0 = 1/3$ was seen. The relevant parameters of the
measurement at $T \approx 60 {\rm K}$ and $H = 300 {\rm Oe}$ depicted for
$B_\phi = 0.2 {\rm T}$ in Fig.~4 of Ref.~\cite{trexp2} are
$f = B / B_\phi \approx 0.15$, and, using $\lambda(0) \approx 1400 \AA$,
$\lambda(T) / d \approx 1.6$. Across about a half--decade the experimental
current--voltage characteristics could be described by an effective exponent
$p_{\rm eff} \approx 0.57$, which agrees very well with our result
$p_{\rm eff} \approx 0.55$ for $f = 0.1$ and $\lambda / d = 2$ (note that both
sets of parameters correspond to the same ratio $\lambda / a_0 \approx 0.6$).
For the experiment with much higher irradiation dose, $B_\phi = 2 {\rm T}$,
corresponding to much lower fillings $f \approx 0.015$ and
$\lambda / d \approx 5$, however, the variable--range hopping exponent turned
out to be {\it lower}, namely $p_{\rm eff} \approx 0.39$, \cite{trexp3}
contrary to the expectations from our above calculations. A possible
explanation is that for such low values of $f$, see also Ref.~\cite{trexp1}
where $p_{\rm eff} \approx 1/2$ was measured for $f \approx 0.0065$, one has to
reconsider the above analysis and take very rare fluctuations in the spatial
distributions of rods into account, leading to $p_{\rm eff} = 1/2$, at least
for the non--interacting case \cite{nelvin}.

Finally, we briefly address the case of samples with finite thickness $L$. At
the very lowest currents $J \rightarrow 0$, the current--voltage
characteristics will eventually become Ohmic, \cite{nelvin} namely when
$L \approx Z^*$, cf. Eq.~(\ref{longsz}); for the power--law density of states
(\ref{gapexp}) studied above the typical longitudinal extent of a superkink is
\begin{equation}
	Z^*(J) = (2 E_{\rm K} / d) (c / \phi_0 J) \quad ,
 \label{pllgsz}
\end{equation}
and hence the crossover current scale to Ohmic resistance becomes
\begin{equation}
	J_L = (2 E_{\rm K} / d) (c / \phi_0 L) \quad .
 \label{plcrcr}
\end{equation}
Substituting this into Eq.~(\ref{pldelf}), we find
\begin{equation}
	\delta F_{\rm SK}^*(L) / T = (L / L_0)^p
 \label{pldelu}
\end{equation}
with the characteristic length scale
\begin{equation}
	L_0 \approx \kappa^{1/(s+1)} (d / 2 E_{\rm K})^D T^{1/p} \quad .
 \label{pllesc}
\end{equation}
The ensuing resistivity law
\begin{equation}
	\rho = \rho_0 \exp \left[ - ( L / L_0)^p \right]
 \label{plresl}
\end{equation}
with its unusual thickness--dependence is the direct analogue of the
variable--range hopping conductivity formula for disordered semiconductors,
\cite{cgapth} generalizing Mott's law to the case of long--range interactions.
The latter is recovered when $s \rightarrow 0$ and hence in $D = 2$ dimensions
$p = p_0 = 1/3$ and $L_0 = g(\mu) (d / 2 E_{\rm K})^2 T^3$ \cite{nelvin}.


\section{Summary and Discussion}
 \label{sumdis}

We have studied spatial correlations and the interacting density of states of
flux lines pinned to columnar defects in the Bose glass phase \cite{nelvin}.
Our procedure was to map the vortex problem onto an equivalent two--dimensional
disordered boson system at zero temperature, \cite{nelson,lyutov,nelvin} and
examine the latter using an established Monte Carlo simulation algorithm
\cite{cgapth,leesim,recsim,gapsim}. We expect our simplified Hamiltonian to
apply for $B < B^*$ [Eq.~(\ref{crossb})] and ``low'' temperatures $T < T_1$
[Eq.~(\ref{crosst})], which may in fact extend throughout a large fraction of
the Bose glass phase, up to temperatures very near to the irreversibility line.
The resulting distribution of pinning energies $g(\epsilon)$, which explicitly
takes the possibly long--range interactions into account, was then used to
calculate the vortex transport characteristics in the variable--range hopping
regime at low currents $J < J_1$ [Eq.~(\ref{loopcr})], by generalizing the
optimization procedure of Ref.~\cite{nelvin}.

For any $\lambda \geq a_0$ and low fillings $f \leq 0.3$, we have found that
the flux lines form a highly correlated structure, the corresponding $S(q)$
displaying a marked, but broad peak at $q a_0 \approx 2 \pi$. This spatial
configuration may be understood as a highly distorted triangular lattice, with
a large number of topological defects in the vortex coordination shells imposed
by the underlying random distribution of columnar pinning sites. The peak in
$S(q)$ is destroyed for larger fillings, and/or for weaker interactions
$\lambda < a_0$. As for the Coulomb glass, \cite{leesim} a peculiar spatial
clustering of those sites with pinning energies near the chemical potential
$\mu$ was found, the empty low energy sites and filled high energy sites
occupying roughly complementary regions of space. As opposed to the case of
localized charge carriers in disordered semiconductors, \cite{cgapth} the above
spatial structures can be directly investigated in experiment, and the first
such measurements have already been reported \cite{stmexp,decexp}. Quite
unambiguously, highly correlated vortex distributions were found, and thus
the flux line repulsions need to be accounted for \cite{decexp}. Moreover, we
have demonstrated how measured data on columnar pin and vortex positions may be
utilized to actually predict the corresponding density of states and estimate
the current--voltage characteristics \cite{letter}.

In the distribution of pinning energies (single--particle density of states),
for $\lambda \geq a_0$ the interactions lead to the formation of a remarkably
prominent and persisting ``Coulomb'' gap near the chemical potential $\mu$
separating the occupied and empty states. This soft pseudo--gap can at least
approximately be described by a power law (\ref{gapexp}), characterizing the
depletion of states upon approaching $\mu$. With increasing filling $f$ for
fixed interaction range $\lambda$, the gap exponent becomes smaller and the
pseudo--gap eventually closes as the vortices have to accomodate with the
underlying randomness, thus losing their spatial correlations. With our
simulations, we found effective gap exponents $s_{\rm eff} \approx 1$ for
$\lambda \approx d$ up to values $s_{\rm eff} \approx 3$ for
$\lambda \rightarrow \infty$ and $f \ll 1$, which demonstrates that correlation
effects are in fact much stronger than predicted by the qualitative mean--field
estimate (\ref{loggap}).

These equilibrium results were used as a basis to discuss both the half--loop
nucleation region at intermediate currents, as well as the variable--range
hopping transport regime at very low currents, where the vortices move by
forming double--superkinks to favorable, possibly distant sites \cite{nelvin}.
For the former case, we found that $p_{\rm eff} = 1$ as in the non--interacting
situation. In the variable--range hopping regime, however, the depletion of
low--energy states according to Eq.~(\ref{gapexp}) implies a change of the
effective transport exponent $p_{\rm eff}$ to higher values, rendering the
collective pinning of vortices to columnar defects much more effective when
$\lambda \geq a_0$. For $\lambda = d$ we have found $p_{\rm eff} \approx 1/2$,
while for very long--range repulsion $\lambda \rightarrow \infty$ values up to
$p_{\rm eff} \approx 2/3$ may be reached, similar to the scaling analysis by
Fisher, Tokuyasu, and Young \cite{ftyexp}. In at least one measurement an
effective exponent $p_{\rm eff} \approx 0.57$ was found for parameter values
well in accord with our simulations \cite{trexp2}.

Therefore, we may speculate that the remarkable correlation effects reported in
this paper, which are induced by the long--range intervortex repulsions, could
in fact be utilized for designing future superconducting materials in order to
reduce dissipative flux transport as much as possible. For example, one could
specifically ``tailor'' samples to obtain large values of $\lambda / a_0$. The
remarkably strong pinning predicted for splayed columnar defects \cite{splays}
might be even further enhanced by interactions.


\acknowledgments

We are indebted to E.~Frey, who provided us with the subroutine performing the
Ewald summation for logarithmic interactions, and to H.~Dai, who prepared the
Voronoi plots of Fig.~\ref{vrlami}.
We would like to thank H.~Dai and C.M.~Lieber for sharing their data with us
prior to publication.
We benefitted from discussions with A.L.~Efros, D.S.~Fisher, E.~Frey, T.~Hwa,
P.~Le Doussal, and V.M.~Vinokur.
This research was supported by the National Science Foundation, in part by the
MRSEC Program through Grant DMR-9400396, and through Grant DMR-9417047.
U.C.T. acknowledges support from the Deutsche Forschungsgemeinschaft (DFG)
under Contracts Ta.~177/1-1,2.




\begin{table}
\setdec 0.00
\caption{Boson analogy applied to vortex transport.}
\medskip

\begin{tabular}{lcccccccc}
Charged bosons & Mass & $\hbar$ & $\hbar / T$ & Pair potential & Charge &
Electric field & Current \\
\tableline
Superconducting vortices & ${\tilde \epsilon}_1$ & $T$ & $L$ & $2 \epsilon_0
K_0(r/\lambda)$ & $\phi_0$ & ${\bf {\hat z}} \times {\bf J}/c$ & ${\cal E}$ \\
\end{tabular}
 \label{bosmap}
\end{table}


\begin{figure}

FIG.~\ref{colpin}. Columnar defects and pinned flux lines, with filling
	 fraction $f < 1$.

FIG.~\ref{phadia}. Schematic phase diagram of flux lines interacting with
	 columnar defects that are aligned along the magnetic field
	 direction.

FIG.~\ref{vrkink}. Double--superkink configuration, which is the most important
	 excitation relevant for the variable--range hopping transport regime
	 of vortices.

FIG.~\ref{curvol}. Schematic current--voltage characteristics in the Bose glass
	 phase.

FIG.~\ref{xyconf}. Spatial positions of unoccupied columnar pins (open circles)
	and columnar pins occupied by magnetic flux lines (filled circles), as
	obtained from typical simulations (with randomly chosen initial vortex
	configuration) using an identical set of underlying defect coordinates
	($N_{\rm D} = 400$). The parameters are: $w / 2 \epsilon_0 = 0.1$,
	$\lambda \rightarrow \infty$ (i.e.: logarithmic interactions), and
	(a) $f = 0.1$, (b) $f = 0.2$, and (c) $f = 0.4$.

FIG.~\ref{sqconf}. Vortex structure function $S(q)$, obtained by averaging over
	the initial and final states of a series of $k = 40$ different defect
	configurations with $N_{\rm D} = 400$, $w / 2 \epsilon_0 = 0.1$,
	$\lambda / d = 5$, and (a) $f = 0.1$, (b) $f = 0.2$.
	$S(q)$ for the initial flux line distributions is the long--dashed line
        for randomly chosen positions, and the dashed line for the distorted
	triangular lattice; the final vortex distributions are depicted as the
	dot--dashed line for the configurations stemming from random initial
	states, and the solid line originating in the distorted triangular
	lattice positions. Notice the shift in the peak position of the dashed
	curves in (a) and (b).

FIG.~\ref{sqfill}. Dependence of the vortex structure function $S(q)$ on the
	filling $f$. The results were obtained by averaging over $k = 40$
	different defect configurations (with $N_{\rm D} = 400$ and random
	initial occupations), where $w / 2 \epsilon_0 = 0.1$ and
	$\lambda / d = 2$; solid: $f = 0.1$, dot--dashed: $f = 0.2$, dashed:
	$f = 0.4$, and long--dashed: $f = 0.8$. A broad peak at
	$q a_0 \approx 2 \pi$ is found only for $f \leq 0.4$.

FIG.~\ref{vrlami}. Voronoi analyses (Delaunay triangulations) of the vortex
	positions shown in Fig.~5; i.e.: $N_{\rm D} = 400$,
	$w / 2 \epsilon_0 = 0.1$, $\lambda \rightarrow \infty$, and
	(a) $f = 0.1$, (b) $f = 0.2$, (c) $f = 0.4$.
	The vortices with sixfold coordination are depicted as full circles,
	while topological defects with coordination number $N_c=4$ are shown as
        open circles, $N_c=5$ as open diamonds, $N_c=7$ as encircled asterisks,
	and $N_c=8$ as encircled crosses.
	The histograms depict the relative abundancy $h(N_c)$ for the occurence
	of the coordination numbers $N_c = 4, 5, 6, 7, 8$, respectively.

FIG.~\ref{cllami}. Spatial clustering of both occupied (filled circles) and
	empty (open circles) sites with energies near the chemical potential,
	for the final configuration of Fig.~5(c), i.e.: $N_{\rm D} = 400$,
	$f = 0.4$, $w / 2 \epsilon_0 = 0.1$, and $\lambda \rightarrow \infty$.
	The small symbols depict the complete distribution of vortices and
	pins, while the large circles correspond to those sites whose energies
	$\epsilon_i$ lie in the range
	(a) $| \epsilon_i - \mu | / 2 \epsilon_0 \leq 0.6$, and
	(b) $| \epsilon_i - \mu | / 2 \epsilon_0 \leq 1.2$, respectively.

FIG.~\ref{dslami}. Normalized distribution of pinning energies
	$G(E') = 2 \epsilon_0 d^2 g(\epsilon)$ as function of the
	single--particle energies $E' = \epsilon' / 2 \epsilon_0$, averaged
	over $k = 2 \times 40$ runs with $N_{\rm D} = 400$,
	$w / 2 \epsilon_0 = 0.1$, $\lambda \rightarrow \infty$, and
	(a) $f = 0.2$, (b) $f = 0.4$;
	$\mu' / 2 \epsilon_0 \approx -1.3$ (marked by the arrow) in both cases.
	Open circles: results of the simulations starting from random initial
	configurations, filled triangles: site--energies from runs where a
	distorted triangular lattice was used as initial state.

FIG.~\ref{ldlami}. Double--logarithmic plots of the normalized distribution of
	pinning energies $G(E) = 2 \epsilon_0 d^2 g(\epsilon)$vs.
	$| E - E_\mu | = | \epsilon - \mu | / 2 \epsilon_0$, averaged over
	$k = 2 \times 40$ runs with $N_{\rm D} = 400$,
	$w / 2 \epsilon_0 = 0.1$, $\lambda \rightarrow \infty$, and
	(a) $f = 0.1$, (b) $f = 0.4$.
	Circles and triangles refer to the choices of random positions or a
	distorted triangular lattice as initial configurations, as in the
	previous figure. Filled and open symbols refer to single--particle
	energies in the occupied ($n_i = 1$) and empty ($n_i = 0$)
	pseudo--bands, respectively.

FIG.~\ref{dslam5}. Normalized distribution of pinning energies
	$G(E) = d^2 g(\epsilon)$ as function of the single--particle energies
	$E = \epsilon / 2 \epsilon_0$, averaged over $k = 2 \times 40$ runs
	with $N_{\rm D} = 400$, $w / 2 \epsilon_0 = 0.1$, $\lambda / d = 5$,
	and (a) $f = 0.1$: $\mu / 2 \epsilon_0 \approx 10.9$,
	(b) $f = 0.4$: $\mu / 2 \epsilon_0 \approx 46.9$,
	(c) $f = 0.8$: $\mu / 2 \epsilon_0 \approx 96.8$.
	The symbols have the same meaning as in Fig.~10.

FIG.~\ref{dslam2}. As in Fig.~12, but for $\lambda / d = 2$ and
	(a) $f = 0.1$: $\mu / 2 \epsilon_0 \approx 1.74$,
	(b) $f = 0.4$: $\mu / 2 \epsilon_0 \approx 9.07$.

FIG.~\ref{dslam1}. As in Fig.~12, but for $\lambda / d = 1$ and
	(a) $f = 0.2$: $\mu / 2 \epsilon_0 \approx 0.72$,
	(b) $f = 0.8$: $\mu / 2 \epsilon_0 \approx 5.33$.

FIG.~\ref{mulam5}. Chemical potential $E_\mu = \mu / 2 \epsilon_0$ as function
	of the system size ($N_{\rm D} = 100$, $200$, $400$, $800$), averaged
	over $k = 16000 / N_{\rm D}$ runs (with random initial vortex
	distributions), for $w / 2 \epsilon_0 = 0.1$ and $\lambda / d = 5$;
	squares: $f = 0.1$, triangles: $f = 0.2$, circles: $f = 0.3$, inverted
	triangles: $f = 0.4$, diamonds: $f = 0.5$. The open symbols plotted at
	$1/N_{\rm D} = 0$ represent the corresponding results for
	${\tilde \mu} = \mu'(N_{\rm D} = 400) + 2 \pi f (\lambda / d)^2 $
	obtained with the chemical potential results $\mu'$ for the purely
	logarithmic potential ($\lambda \rightarrow \infty$).

FIG.~\ref{ivfill}. Double--logarithmic plots of the exponential factor
	$R^*(J) / d \propto \delta F_{\rm SK}^*(J)$ for variable--range hopping
	vs. the reduced current $j = J \phi_0 d / 2 \epsilon_0 c$, derived from
	the numerically obtained density of states, averaged over $k = 40$ runs
	with $N_{\rm D} = 400$, random initial configurations, and
	$w / 2 \epsilon_0 = 0.1$, for different interaction ranges: diamonds:
	$\lambda \rightarrow 0$, circles: $\lambda / d = 1$, squares:
	$\lambda / d = 2$, triangles: $\lambda / d = 5$; (a) $f = 0.1$,
	(b) $f = 0.4$, and (c) $f = 0.8$.

\end{figure}

\end{multicols}

\begin{references}
\bibitem{kbnote} Throughout this paper, we set $k_{\rm B} = 1$, i.e.:
		temperatures are measured in energy units.
\bibitem{fnote1} In the limit $\lambda \rightarrow \infty$ the chemical
		potential contains an additive term $4 \pi f \epsilon_0
		(\lambda / d)^2$, which diverges when the energies are measured
		in units of $\epsilon_0$, and corresponds to the energy of a
		``neutralizing'' background (interpreting $\phi_0$ as charge);
		\cite{supsol} this contribution should thus be subtracted
		first. Physically, the neutralizing background cooresponds to
		adding the magnetic field energy to the free energy
		(\ref{modelf}).
\bibitem{fnote2} Note that Eq.~(\ref{epsshi}) does not imply that the chemical
		potential will diverge for logarithmic interactions, as the
		overall energy scale was $\epsilon_0 \propto \lambda^{-2}$,
		which cancels the dependence of the additive term on the London
		penetration depth.
\bibitem{fnote3} Had we taken the average over the shifted energy differences
		${\tilde \epsilon}_i = \epsilon_i - \mu$, we would have found
		$g(\mu) = 0$ by construction \cite{cgapth}.
\end{references}
\end{document}